\documentclass[5p]{elsarticle}
\usepackage{lineno,hyperref}
\usepackage{amsmath,amssymb,amsfonts}
\usepackage{algorithmic}
\journal{Computer Physics Communications}
\begin{document}
\begin{frontmatter}

\title{Optimized thread-block arrangement in a GPU implementation of a linear solver for atmospheric chemistry mechanisms}

\author[1]{Christian Guzman Ruiz}
\author[1]{Mario C. Acosta}
\author[1]{Oriol Jorba}
\author[2]{Eduardo Cesar Galobardes}
\author[1,3]{Matthew Dawson}
\author[1]{Guillermo Oyarzun}
\author[1]{Carlos Pérez García-Pando}
\author[1]{Kim Serradell}
\address[1]{Barcelona SuperComputing Center, Barcelona, Spain}
\address[2]{Universitat Autonoma de Barcelona, Bellaterra, Spain}
\address[3]{Atmospheric Chemistry Observations and Modeling Laboratory National Center for Atmospheric Research}

\begin{abstract}
Earth system models (ESM) demand significant hardware resources and energy consumption to solve atmospheric chemistry processes. Recent studies have shown improved performance from running these models on GPU accelerators. Nonetheless, there is room for improvement in exploiting even more GPU resources.

This study proposes an optimized distribution of the chemical solver's computational load on the GPU, named Block-cells. Additionally, we evaluate different configurations for distributing the computational load in an NVIDIA GPU. 

We use the linear solver from the Chemistry Across Multiple Phases (CAMP) framework as our test bed. An intermediate-complexity chemical mechanism under typical atmospheric conditions is used. Results demonstrate a 35$\times$ speedup compared to the single-CPU thread reference case. Even using the full resources of the node (40 physical cores) on the reference case, the Block-cells version outperforms them by 50\%. The Block-cells approach shows promise in alleviating the computational burden of chemical solvers on GPU architectures.
\end{abstract}

%\begin{highlights}
%Maximum 125 characters for highlight, limit of 3 to 5 highlights
%\item Integration of a Biconjugate Gradient algorithm as GPU linear solver for chemistry in Earth System Models
%\item Development of a strategy to eliminate communication between GPU thread blocks (Block-cells approach)
%\item Performance assessment of different thread block arrangements 
%\item Up to 35$\times$ linear solver speedup over the base single-thread MPI configuration, and up to  50\% faster over the multi-core MPI configuration (40 cores)
%\end{highlights}


\end{frontmatter}

\section{Introduction}

Earth system models (ESM) aim to represent all relevant interactions of the Earth system components (i.e., atmosphere, ocean, sea ice, land surface, biosphere, ice sheets). Much of their complexity and computational burden can be attributed to the atmospheric component, which solves the atmosphere’s physical, chemical, and radiative processes \cite{zhang_chemical_2011} \cite{jacobson_fundamentals_2005}. Large computational costs and vast amounts of data are typical of ESMs. For this reason, the workload is usually parallelized using the domain decomposition strategy (each computational task solves a fraction of the domain of study with an exchange of information among them) in a High-Performance Computing (HPC) environment. Traditionally, ESMs have been designed for CPU execution. With the advancement of HPC technology, the Earth Sciences community is trying to run the models on GPU systems, which usually leads to performance improvements over the traditional CPU--based computation \cite{navarro_survey_2014}. However, refactoring the code efficiently to a different parallel computing paradigm is a complex and long-term task, particularly for models involving hundreds of thousands of lines of code like ESMs \cite{alexander_software_2015}. Hence, most optimization work on ESMs generally focuses on specific components of the model rather than on the entire system, preserving the CPU version for portability and leveraging the computational power of accelerators for particular parts of the code \cite{kelly_gpu_2010}.

Among ESM components, chemistry stands out for its computational burden, which can account for up to 90\% of the total execution time of the model \cite{christou_earth_2016}. The solution of a chemical mechanism (a set of chemical reactions describing the fate of hundreds or thousands of chemical species with a wide range of lifetimes) can be the most expensive part of the problem. The temporal evolution of the concentration of each chemical species can be described by a set of ordinary differential equations (ODE). The resulting ODE system is mathematically \textit{stiff} \cite{shampine_users_1979}, and special care needs to be exercised in the choice of the numerical integration scheme \cite{zhang_chemical_2011}. Other factors that can affect the computational cost of the solution of an ODE system are the desired accuracy and the hardware architecture used.

Historically, chemical solvers have been designed for single-thread execution. For example, this approach is employed by the Kinetic PreProcessor \cite{damian_kinetic_2002}, which has been widely used to generate gas-phase mechanism code with several ODE solver options. However, recent studies have shown performance improvements after porting chemical solvers to GPUs. Alvanos and Christoudias \cite{alvanos_accelerating_2019} deployed a GPU version of KPP in the EMAC ESM, reporting speedups between 1.19$\times$ up to 1.9$\times$ for a node-to-node real-world production scenario. Sun et al. \cite{sun_computational_2018} obtained an 11.7$\times$ speedup porting the Rosenbrock solver used in the CAMS4-chem model to GPUs compared to single-process kernel execution.

Nevertheless, applying strategies explicitly developed for GPU computing has the potential for further performance gains. For example, the Runge-Kutta-Chebyshev (RKC) and Runge-Kutta-Cash-Karp method (RKCK) algorithms for slightly and moderately stiff chemical kinetics can achieve up to a 59$\times$ speedup relative to a single-thread CPU application \cite{niemeyer_accelerating_2014}. In another work, Guzman et al. \cite{Guzman2021} have shown the benefits of exploiting the parallelization of internal routines within a chemical system without translating the entire chemical solver to new parallel paradigms like GPU.

To the best of the authors’ knowledge, most of the current GPU solutions for chemistry solvers follow a similar parallelization strategy. Specifically, each computational process (i.e., GPU thread) solves the workload of a cell. A cell is the smallest unit obtained by discretizing the domain of study (e.g., the atmosphere in atmospheric models). Each cell describes the state of the atmosphere in that specific domain region (including variables such as chemical species concentrations, temperature, pressure, etc.). In each cell, the prognostic equations describing the evolution of state variables are solved through multiple algebraic operations over the concentrations array. During solving, many calculations related to specific species are independent. This means that they can be computed in parallel.

In this work, we present a new strategy to improve the computational load distribution of a GPU chemical solver. Namely, each GPU thread computes the workload of a species concentration within a cell, resulting in a set of processes whose number equals the number of cells times the number of chemical species. We name this approach Block-cells. In this way, we increase the distribution or parallelization of the load beyond the work assigned to one cell per thread. This leads to significantly improved exploitation of the high bandwidth capacity of the GPU. Applying this change requires extra development work to efficiently communicate data dependencies between species in the same cell, as well as transforming concentration array loops into parallel tasks. However, we demonstrate the rewards of this effort in terms of improved performance and present a novel approach that should encourage the community to seriously consider porting complex chemical solvers to accelerators. 

We demonstrate the benefit of this methodology using the Chemistry Across Multiple Phases (CAMP) framework \cite{dawson_chemistry_2022}. CAMP provides a flexible environment for integrating chemical mechanism solvers into host models of various complexity (e.g., from box models to ESMs). We have extended the previous work of Guzman et al.\cite{Guzman2021} with the implementation in CAMP of a linear solver optimized for use on GPUs: a Biconjugate Gradient (BCG) linear solver \cite{oyarzun_mpi-cuda_2014}. We compare the default CPU--based KLU linear solver \cite{davis_algorithm_2010} available in CAMP with the new GPU--based linear solver to evaluate the Block-cells approach.  

The application context used in this work (CAMP library and the KLU and BCG linear solvers) is described in Section \ref{Background}. Section \ref{Implementations} introduces the new Block-cells approach. In Section \ref{Test environment}, we present the hardware and software configuration for the tests performed. Results are discussed in Section \ref{Results}. Finally, Section \ref{conclusions} presents concluding remarks and future work.

\section{Background: Application context} \label{Background}

\subsection{Atmospheric chemistry models}

An atmospheric model is a mathematical representation of the spatial and temporal distribution of state variables in the atmosphere. The computational domain is either global or regional, composed of cells representing a fractional volume of the atmosphere \cite{steyn_regional_2012}. If chemistry processes are considered, such models are known as atmospheric chemistry models or chemical transport models \cite{seinfeld_atmospheric_1998}. The physicochemical processes considered in such a model are the emissions of inert or reactive chemical species from anthropogenic or natural sources, the transport by advection in the direction of the wind and lateral and vertical diffusion, the photochemical transformations in the atmosphere, and the dry and wet deposition towards the surface. Our experiments are applied to multi-phase chemical kinetics as they can account for most of the execution time \cite{christou_earth_2016}. Note that our implementation is independent of the type of chemical reactions considered (e.g., unimolecular, bimolecular, termolecular, or photochemical). The cells can be treated independently by the ODE solver, providing a suitable environment for exploiting different parallelization strategies.

Traditionally, atmospheric models are parallelized following a domain decomposition approach. This approach yields parallelization without the development effort of other paradigms. Each parallel process is often configured to solve thousands of cells, as the number of cells in the domain dramatically outnumbers the number of parallel processes. As these models have been historically developed for CPU--based architectures, MPI and OpenMP parallelization paradigms are commonly used. However, computation can be further parallelized using other parallel architectures, such as GPUs, in combination with MPI parallelization. Several studies have reported positive results applying this combination in physical models \cite{yamagishi_gpu_2016} \cite{borrell_heterogeneous_2020}. This work presents a GPU--based alternative to an MPI CPU--based reference as a starting point for the future development of a joint MPI--GPU coupled atmosphere model.

\subsection{Chemical mechanism solver} \label{CAMP}

We use the Chemistry Across Multiple Phases (CAMP) framework \cite{dawson_chemistry_2022} as our test bed. CAMP is a novel framework permitting run-time configuration of chemical mechanisms for mixed gas- and aerosol-phase chemical systems.

It also allows an abstract, non-fixed representation of aerosols that can be configured at run-time. It has been developed to minimize hard-coded components in the code, facilitating new developments such as integrating GPU routines presented in this work. In addition, it is designed to allow gas and aerosol reactions to be solved kinetically in a unified system simultaneously. Hence, the introduction of new solvers can lead to the optimization of both gas-phase and aerosol chemistry. This may lead to better overall performance of a host model (e.g., an Eulerian atmospheric model), as solving the chemical system can account for a significant fraction of the total computational burden of an atmospheric model. Furthermore, CAMP builds as an external software library with a well-defined API, facilitating its rapid integration in host models with diverse complexity. 

CAMP is designed to use external ODE solvers to solve the chemistry time-dependent equation ( $y' = f(t, y)$ ). The default version of CAMP is coupled to the external CVODE solver of the SUNDIALS package using backward differentiation formulas (BDF) and Newton iteration \cite{cohen_cvode_1996} \cite{serban_cvodes_2008}. This algorithm is suitable for mathematically stiff systems. The variable-order, variable time-step CVODE solver with time-step error control provides accurate solutions. Thus, it was chosen as the initial solver option for CAMP \cite{dawson_chemistry_2022}. The BDF algorithm requires the solution of a linear system at each integration step. CAMP is configured by default to use the KLU Sparse solver \cite{davis_algorithm_2010}. The sparse structure avoids storing zero values in the Jacobian matrix, which are common for chemical systems \cite{xu_sparse_2016}.

\begin{figure}[!t]
  \centering
\includegraphics[width=1\linewidth]{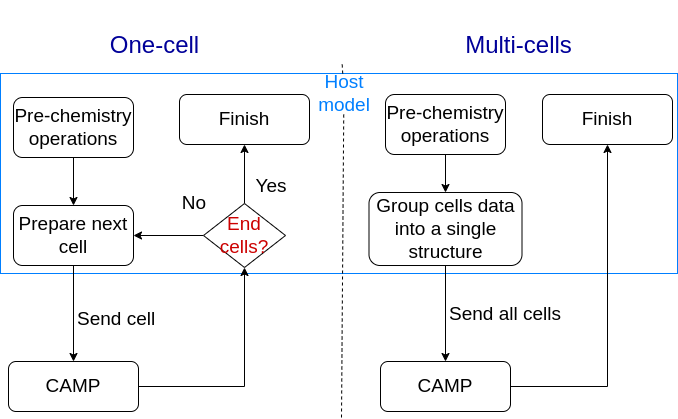}
  \caption{Workflow diagram of the One-cell (left) and Multi-cells (right) solving strategies.}
  \label{multi_one}
\end{figure}

CAMP can easily be implemented in an atmospheric host model following one of two approaches: One-cell or Multi-cells \cite{Guzman2021}. Figure \ref{multi_one} shows a schematic workflow of the two approaches. One-cell corresponds to most models’ classical implementation of chemical mechanism solvers. In this configuration, CAMP is employed to advance the concentrations of the chemical species in each model grid cell sequentially in time. The overall rate of change for each species $y_i$ and reaction $j$ at any given time is thus,

\[ f_i \equiv \frac{dy_i}{dt} = \sum_j \left(\frac{dy_i}{dt}\right)_j, \]

In contrast, the Multi-cells approach takes advantage of the flexibility of CAMP by grouping multiple cells in batches, and solving the chemical mechanism for each batch of cells simultaneously. As an example, the equation is updated as follows:

\[ f_i \equiv \frac{dy_{ik}}{dt} = \sum_j \left(\frac{dy_{ik}}{dt}\right)_j \]

where $y_ik$ refers to the species $y_i$ from cell $k$.

With this approach, there is no need to loop over cells or re-initialize ODE solver parameters and data structures for each cell, and solver iterations could be reduced by a factor of 10$^{4}$, resulting in up to a 14$\times$ speedup over the One-cell approach \cite{Guzman2021}. Moreover, the Multi-cells approach maximizes the amount of information passed to the solver simultaneously, making it possible to explore massive parallelism. However, Multi-cells could lead to less accuracy since it solves a single enormous structure with all the ODE equations coming from each cell instead of solving them separately. For this reason, we present a different approach, named Block-cells, explained in section \ref{Implementations}.

Although a hybrid MPI/OpenMP version and a performance comparison between MPI processes and OpenMP threads would be interesting, it's important to consider that this study aimed not only to obtain a GPU-portable implementation that enhances the parallel CPU version but also to maintain absolute compatibility of the chemistry component with the MPI-based atmospheric models in which CAMP is included (such as the MONARCH model \cite{dawson_chemistry_2022}). These models utilize MPI for parallelization and domain decomposition, implying that using OpenMP for only a part of the complete model may not be efficient. Additionally, there is evidence from past studies that similar models achieved similar efficiency between an MPI implementation and OpenMP \cite{szustak_toward_2020} \cite{armstrong_quantifying_2000}. However, we reserve a definitive conclusion regarding these assumptions for future work.

\section{Block-cells implementation} \label{Implementations}

In this Section, we present our methodology to optimize the computational load of a linear solver using GPUs, namely the Block-cells approach. As a first step, we have coupled a GPU--based linear solver in CAMP to analyze the performance of different GPU--based parallelization strategies: One-cell, Multi-cells and Block-cells. As a second step, we evaluated different kernel configurations of the Block-cells approach.

\subsection{Coupling a GPU BCG linear solver in CAMP} \label{BCG}

Originally, CAMP’s only linear solving option was the CPU--based KLU sparse linear solver. However, to leverage the benefits of GPU architecture, we have coupled CAMP to a sparse CUDA version of the Biconjugate Gradient (BCG) algorithm developed at the Barcelona Supercomputing Center \cite{oyarzun_mpi-cuda_2014}. The BCG algorithm was chosen because multiple studies have found it performs better than other Conjugate Gradient (CG) methods \cite{babaoglu_application_2003}. Moreover, it is designed specifically for use on GPUs, making it a better candidate for this application than simply translating a CPU-focused algorithm like KLU.

We applied the BCG solver to the One-cell and Multi-cells implementations. In the One-cell version, each call made to the GPU passes the state of a single cell. As a cell comprises hundreds of species, this under-utilizes the millions of threads available on modern GPUs. The Multi-cells approach, in contrast, gathers all the cells into a single data structure before any GPU computation, preparing all the data for parallel execution, resulting in a promising solution to the problem of adequately exploiting the capacity of the GPU. In this approach, the GPU can simultaneously compute the states of thousands of cells.

\begin{figure}[!t]
  \centering
  \includegraphics[width=0.7\linewidth]{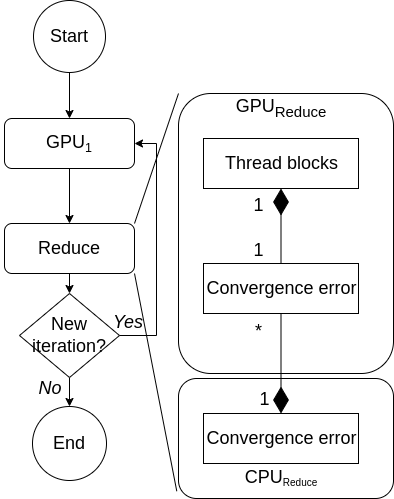}
      \caption{Diagram of BCG Multi-cells interactions between CPU and GPU (excluding data transfers). $GPU_1$ contains all the BCG operations in the GPU except for $Reduce$ (e.g., functions like dot vector, SPMV, etc.). The $Reduce$ computation is divided between CPU and GPU. Each thread block performs a reduction operation on the GPU, resulting in a convergence error for each block. Then, another reduction is performed over these errors on the CPU side. The final value of the reduction is used to check for convergence, i.e., if the algorithm needs to iterate again or finish.}
  \label{UMLReduce}
\end{figure}

However, the Multi-cells implementation requires an extra reduction operation on the CPU. This operation involves summing each element of an array to obtain a final single value, which determines if the BCG performs another iteration or finishes. Figure \ref{UMLReduce} illustrates how, in the Multi-cells approach, data is transferred to the CPU during each solving iteration to perform the reduction and convergence checking. This data reduction can account for more than 50\% of the total execution time \cite{xiao_inter-block_2010}.

\begin{figure}[!t]
  \centering
  \includegraphics[width=\linewidth]{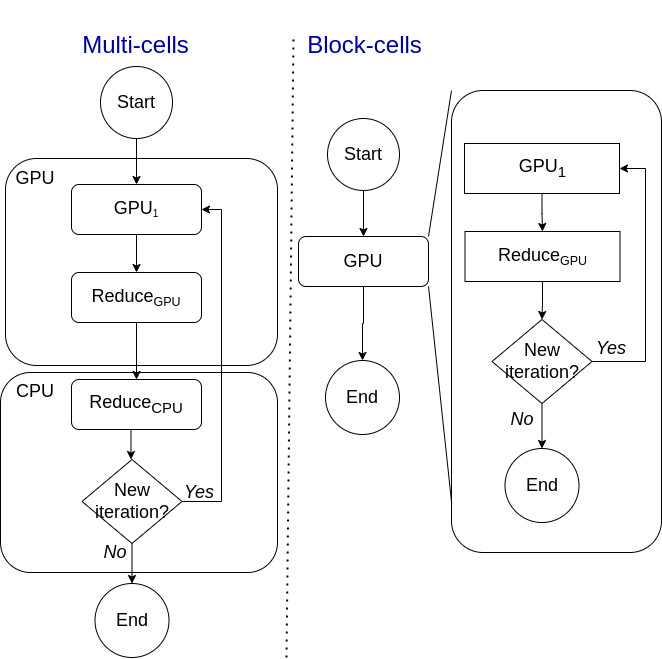}
      \caption{Diagram of Multi-cells and Block-cells interactions between CPU and GPU (excluding data transfers).}
  \label{UMLMultiBlock}
\end{figure}

This extra reduction is necessary with the Multi-cells approach, as convergence must be evaluated for the entire system comprising all cells. However, the full system can also be treated as multiple independent systems of cells as in the One-cell implementation and form the basis for a new parallel implementation. The computational load for the system’s cells can be distributed across blocks, following a CUDA thread block distribution \cite{ghorpade_gpgpu_2012}. This way, the CPU reduction operation can be avoided, encapsulating the BCG operations in a single kernel call. This implementation is called the Block-cells approach, illustrated in Figure \ref{UMLMultiBlock}.

However, to avoid communication between GPU blocks, the chemical species concentrations to calculate should be equal to or less than the maximum block size. In modern GPUs, this block size is usually 1024 threads. Only in rare cases do ESMs use mechanisms with more than 1024 species. Often, chemical mechanisms comprise less than two hundred species. Therefore, each block can accommodate one or more cells. For example, a mechanism with 100 species can be solved with up to 10 cells in a block. In this work, we evaluate the performance of various configurations of block size. 

\subsection{Block-cells kernel configurations} \label{BCG conf}

In the following, we refer to the number of species in a cell as the “cell size.”

In Block-cells (1), the number of threads per block corresponds to the number of species concentrations in a cell. The number of blocks corresponds directly to the number of cells.

In Block-cells (N), we calculate the maximum number of cells per block without partitioning any cell. It is calculated by dividing the block size by the cell size and rounding down the result. The residual of the division is calculated in a separate kernel. For example, in an experiment with 11 cells, 100 chemical species, and 1024 threads per block, the GPU would run a kernel with 1000 threads corresponding to 10 cells and another kernel of 100 threads for the last cell.

In Block-cells (2) and Block-cells (3), we test 2 and 3 cells per block, respectively. This range is in the middle of Block-cells (1) and Block-cells (N), allowing us to see how the performance varies with block size and to find the optimal value. Table \ref{GPU1} summarizes the cells-per-block and threads-per-block configurations.

\begin{table}
\caption{Cells per block and threads per block of Multi-cells, Block-cells (1), and Block-cells (N). $float(Cells/block)$ indicates that part of a cell can be computed in a block while another block would compute the rest. $int(Cells/block)$ indicates the opposite; each cell should be computed by a single block. Chemical $Species$ depends on the chemical configuration used.}
\label{GPU1}
\resizebox{\linewidth}{!}{%
\begin{tabular}{llll}
\hline
Case & Cells/block & Threads/block\\
\hline
Multi-cells        & float(Cells/block)             & Maximum  \\
Block-cells (1)    & 1               & Species \\
Block-cells (2)    & 2               & Species \\
Block-cells (3)    & 3               & Species \\
Block-cells (N)    & int(Cells/block)               & int(Cells/block)*Species\\
\hline
\end{tabular}%
}
\end{table}

The reduction operation in the GPU uses a shared memory array, whose length is always set to a power of two, following recommendations from NVIDIA developers intended to improve the efficiency of reduction operations \cite{harris_optimizing_nodate}. Therefore, if the number of threads per block is not a power of two, the shared memory is set to the next power of two (e.g., for 100 threads per block, the next power of two is 128). Table \ref{GPU2} lists the shared memory configurations for the Multi-cells and both Block-cells (1) and (N) implementations.

\begin{table}
\caption{Shared memory of Multi-cells, Block-cells (1), and Block-cells (N).}
\label{GPU2}
\resizebox{\linewidth}{!}{%
\begin{tabular}{llll}
\hline
Case & Shared memory \\
\hline
Multi-cells        & Maximum          \\
Block-cells (1 to N)    & NextPowerOfTwo(Threads/block)           \\
\hline
\end{tabular}%
}
\end{table}

Both Block-cells (N) and Multi-cells solve a system composed of multiple cells. These require fewer solver iterations than Block-cells (1), as is the case for the Multi-cells and One-cell configurations of the CPU solver \cite{Guzman2021}.

The number of solver iterations is measured differently between the CPU and GPU Block-cells cases. The iterations for the CPU One-cell case correspond to the sum of iterations for all cells, as they are solved sequentially. However, for the GPU Block-cells cases, the cells are solved simultaneously across multiple threads. Therefore, the effective solver iterations correspond to those performed on the last thread block to finish the algorithm. 

We expect Block-cells (1) to require fewer iterations than Block-cells (N) because solving a system of a single cell should be less complex than solving a system with multiple cells. 

Nevertheless, the Block-cells (N) approach still has the advantage of reducing intermediate-solving parameters. For example, instead of storing an error of convergence for each cell, a single variable can be used for multiple cells. In addition, this approach can reduce the number of idle threads. For example, consider a system of 10 cells and 100 threads per block. By design, threads in CUDA are always launched in groups of 32, called warps. Thus, for Block-cells (1), each cell launches 128 threads, resulting in 28 idle threads per cell and 280 idle threads per block. In contrast, for Block-cells (N), the block is composed of 1000 threads, resulting in 24 idle threads, 256 less than Block-cells (1). In other words, grouping cells can be considered a trade-off between a higher utilization of the GPU resources (specifically, memory storage and threads) and more solver iterations.

\section{Test environment} \label{Test environment}

This Section describes the hardware and software configurations used for implementing and testing the proposed approaches described in Section \ref{Implementations}. 

\subsection{Hardware}

All the tests were performed on the CTE-POWER cluster provided by the Barcelona Supercomputing Center \cite{noauthor_support_nodate}. It provides 4 GPUs per node, making it an ideal cluster for GPU-accelerated applications. Detailed hardware and software specifications are described below.

\begin{itemize}
\item Operating system: Red Hat Enterprise Linux Server 7.5 (Maipo).
\item CPU Compiler: GCC version 7.3.0
\item GPU Compiler: NVCC version 10.1.105
\item 2 x IBM Power9 8335-GTH @ 2.4GHz (3.0GHz on turbo, 20 cores and four threads/core, total 40 physical cores per node and 160 virtual threads using hyper-threading)
\item 512GB of main memory distributed in 16 dimms $\times$ 32GB @ 2666MHz
\item 2 x SSD 1.9TB local storage
\item 2 x 3.2TB NVME
\item 4 x GPU NVIDIA V100 (Volta) with 16GB HBM2.
\item Single Port Mellanox EDR
\item GPFS via one fiber link 10 GBit
\end{itemize}

In addition, we use the NVIDIA Visual Profiler (NVVP) from the CUDA toolkit v11.5.1 to visualize the profiling data of the GPU experiments and assess the performance of the tests under analysis.

\subsection{Experimental setup}

CAMP allows the solving of chemical mechanisms of a wide range of complexity, and can treat a combination of gas and aerosol reactions. Here, we select an intermediate complexity gas and aerosol problem used in \cite{dawson_chemistry_2022}. The gas phase chemistry is the Carbon Bond 2005 (CB05) mechanism \cite{Yarwood2005} with fixed photolysis reaction rate constants during the integration. The mechanism is extended with secondary aerosol production from isoprene using a two-product model approximation. In addition, we add emissions at each time step that shift the species concentrations away from equilibrium. The reader is referred to \cite{dawson_chemistry_2022} for further details of the chemical system solved here. 

The absolute tolerance of the CVODE solver is set to 1.0e$^{-4}$. Any error of accuracy below this level of tolerance is considered negligible. The KLU solver uses the same tolerance. The tolerance of the BCG linear solver is set to 1.0e$^{-30}$. This tolerance corresponds to the lowest level of accepted tolerance in CAMP. CAMP uses this low tolerance level to keep chemistry systems positive-definitive, avoiding negative concentrations produced during the CVODE solving. Any negative value greater than ${-1.0e^{-30}}$ produces an extra iteration in the CVODE solving algorithm. Thus, this tolerance avoids any possible extra iterations the BCG algorithm produces.

We use CAMP as a box model where the number of cells to be solved can be configured from 1 to 10,000 cells. The domain decomposition of some atmospheric models allocates around 40,000 cells per MPI process (distributed as 20$\times$20$\times$100 for axes x, y, and z). The 10,000 cells correspond to levels where the performance results are stabilized, and any greater number of cells gives similar results. 

The timing results are averaged over 720 time steps, with a time-step size of 2 min, representing 24 simulation hours. The profiling metrics obtained through NVVP are from the first time-step.

In addition, we evaluate the performance of Multi-cells with various initial conditions as our implementation depends on using CAMP with Multi-cells. Thus, we design two configurations with different initial conditions among cells. One configuration uses the same initial values for all cells. This is referred to as the \emph{ideal} case. The other uses different initial conditions for each cell, referred to as the \emph{realistic} case.

Specifically, the \emph{realistic} case tries to emulate an atmospheric environment. Each cell is considered to be located at a different altitude in the atmosphere, resulting in different initial conditions. First, the pressure is configured to scale linearly with the number of cells from 1000 to 100 hPa. Second, the chemical reactions of type \emph{emissions} also scale linearly with a rate from 1 to 0. In this way, a cell at 100 hPa has 0 emissions, while a cell at 1000 hPa has the maximum emissions value. Finally, the temperature is calculated from the pressure for dry adiabatic conditions \cite{noauthor_skewt_nodate}. 

Table \ref{table:GPU} shows the various configurations of threads-per-block and shared memory analyzed for the GPU implementations. The Multi-cells case solves 6.6 cells per block, which means that part of a cell is computed in another block. Block-cells (1) solves the same threads-per-block as the number of species per cell. Finally, Block-cells (N) truncates the 6.6 cells of Multi-cells to 6 cells, resulting in 924 threads per block. The shared memory length is always configured to a power of two.

\begin{table}
\caption{GPU kernel configuration of the implementation presented in Section \ref{Implementations} and tested in this work}
\label{table:GPU}
\resizebox{\linewidth}{!}{%
\begin{tabular}{llll}
\hline
Case & Cells/block & Threads/block & Shared memory \\
\hline
Multi-cells        & 6.6             & 1024              & 1024          \\
Block-cells (1)    & 1               & 156               & 256           \\
Block-cells (2)    & 2               & 312               & 512           \\
Block-cells (3)    & 3               & 468               & 512           \\
Block-cells (N)    & 6               & 924               & 1024         \\
\hline
\end{tabular}%
}
\end{table}

\section{Results} \label{Results}

This Section presents and discusses the accuracy, performance metrics, and speedup of the BCG implementations for GPU compared to the default CAMP version based on the KLU solver and One-cell approach. The relative error between the species concentrations using the BCG implementations is below the CVODE tolerance of 0.01\%. Hence, results from the new BCG linear solver implementation are not significantly different from those of the base implementation.

\subsection{NVVP profiling}

The NVVP profiling shows that both Block-cells configurations, (1) to (N), spend a similar amount of time executing various types of instructions (memory dependence, synchronization, etc.). Memory operations account for 50\% of the Block-cells execution time, while for Multi-cells this value is 89\%. Thus, the memory dependence bottleneck is reduced by 39\% going from the Multi-cells to the Block-cells configuration as the synchronization is moved to the GPU. Block-cells spends 35\% of the time performing synchronization tasks, as the synchronizations are now performed on the GPU instead of the CPU. Overall, these metrics indicate a performance improvement of Block-cells with respect to Multi-cells.

Table \ref{table:utilization} shows that all Block-cells configurations have similar computation intensity, which is greater than for Multi-cells. This is because Multi-cells applies multiple kernel calls, one for each vector operation---such as a matrix multiplication, a vector by another vector, or a reduction kernel. Thus, the scheduler concatenates operations between kernels, increasing the kernel synchronization overhead. Memory utilization and bandwidth are higher than Block-cells. However, these metrics decrease linearly from Block-cells (1) to (N), indicating that grouping cells reduces the efficiency of processing memory operations.  

\begin{table}
\caption{Device memory bandwidth and utilization of computation intensity and Memory metrics from the NVVP for 10,000 cells run.}
\label{table:utilization}
\resizebox{\linewidth}{!}{%
\begin{tabular}{lllll}
\hline
Case & Computation intensity & Memory & Bandwidth(GB/s) \\
\hline
Multi-cells & 7\%  & 75\% & 715 \\
Block-cells (1)  & 11\% & 65\% & 597 \\
Block-cells (2) & 11\%  & 65\% & 568\\
Block-cells (3) & 11\% & 55\% & 474\\
Block-cells (N) & 11\% & 45\% & 445\\
\hline
\end{tabular}%
}
\end{table}

Next, we evaluate the NVVP metrics of \emph{Global load efficiency}, \emph{Occupancy} and \emph{Warp execution efficiency}, which are defined as follows \cite{noauthor_profiler_nodate}:
\begin{itemize}
    \item \emph{Global load efficiency} is the ratio of requested global memory load throughput to required global memory load throughput expressed as a percentage;
    
    \item \emph{Occupancy} is the ratio of the average active warps per active cycle to the maximum number of warps supported (a warp in CUDA is a group of 32 threads);
    
    \item \emph{Warp execution efficiency} is the ratio of the average active threads per warp to the maximum number of threads per warp expressed as a percentage.
\end{itemize}

Table \ref{table:eff} demonstrates that both global load and warp efficiencies improve across all Block-cells configurations compared with the Multi-cells approach. However, Block-cells (1) and (2) exhibit higher occupancy compared with (3) and (N). This discrepancy arises because Block-cells (1) and (2) are powers of two, whereas the others are not. Consequently, the GPU architecture, optimized for powers of two, organizes resources more efficiently in these cases.

\begin{table}
\caption{Efficiency and occupancy from the Properties view of NVVP for 10,000 cells run.}
\label{table:eff}
\resizebox{\linewidth}{!}{%
\begin{tabular}{llll}
\hline
Case & Global Load Eff. & Warp Execution Eff. & Occupancy \\
\hline
Multi-cells & 24.2\%   & 35.1\% & 68.6\%          \\
Block-cells (1) & 36.5\% & 75.4\% & 61.5\%     \\
Block-cells (2) & 36.3\% & 75.3\% & 61.5\%     \\
Block-cells (3) & 37\%  & 72\%    & 46.7\%     \\
Block-cells (N)  & 36.8\%  & 75.5\%  & 45.3\%  \\
\hline
\end{tabular}
}
\end{table}

We conducted memory requirement estimations by tallying the memory-allocated arrays of the various methods. The memory required is as follows:

\begin{itemize}
\item KLU One-cell: 18KB
\item KLU Multi-cells: 18 KB per cell
\item BCG Multi-cells and Block-cells (1) to (N): 29KB per cell
\end{itemize}

It is worth noting that the Multi-cell and Block-cell approaches utilize identical arrays; the only distinction lies in how they organize the data. At most, the Multi-cells approach requires two additional variables, each with a length equal to the number of cells, for computing the reduction on the CPU. The increased memory requirements of BCG compared with KLU stem from the BCG requirement of nine additional auxiliary arrays.

\subsection{Speedup}

The One-cell CPU-based KLU solver is 2$\times$ faster than the GPU--based One-cell BCG solver. This is a consequence of the GPU One-cell approach, which launches a kernel and transfers data between CPU and GPU for each cell, resulting in significant overhead. This overhead is sharply reduced in the Multi-cells approach, which performs these operations only once for all cells.

We tested a GPU Multi-cells configuration with the reduction operation on the CPU to quantify the impact of the data transfers. The original configuration transfers one variable from the GPU to the CPU for each GPU block, corresponding to the convergence error explained in Figure \ref{UMLReduce}. The modified configuration transfers the full array of data concentrations. As the original configuration uses 1024 threads per block, the new configuration transfers 1024$\times$ more data. Consequently, the new configuration is 12$\times$ slower for 10,000 cells, indicating that the data transfers are very expensive.

\begin{figure}[!t]
  \centering
  \includegraphics[width=\linewidth]{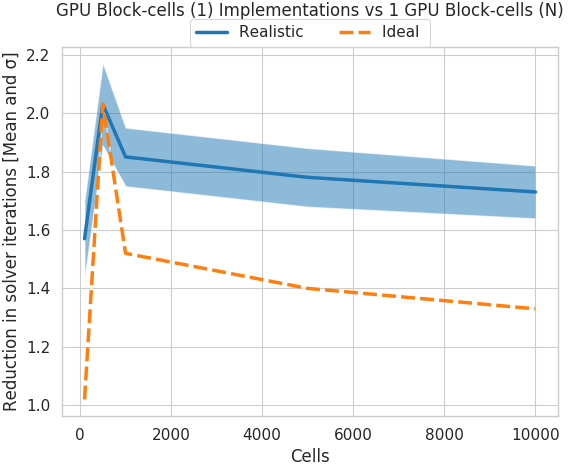}
      \caption{Reduction in the number of solving iterations of Block-cells (1) against Block-cells (N) using \emph{ideal} (orange dashed line) or \emph{realistic} (blue line) initial conditions. This reduction is calculated by dividing the solving iterations of Block-cells (N) by Block-cells (1) and corresponds to the iterations of the last thread block to finish the algorithm. The reduction is averaged over 720 time steps. The blue shade indicates the standard deviation of the reduced iterations for all time steps using \emph{realistic} conditions.}
\label{speedup_iterations_realistic_ideal}
\end{figure}

Figure \ref{speedup_iterations_realistic_ideal} shows that Block-cells (1) iterates less than Block-cells (N) (1.7$\times$ fewer iterations for 10,000 cells and \emph{realistic} conditions). This confirms our expectation that the Block-cells (N) approach generates a more complex system, which takes longer to solve than the slowest cell in Block-cells (1).

Results also indicate that varying initial conditions between the cells increase the performance improvement of Block-cells (1) compared with Block-cells (N), as more iterations are reduced under \emph{realistic} than \emph{ideal} conditions. Moreover, these variations of the initial conditions increase the standard deviation. The plots show a low standard deviation (0.1 for a 1.8$\times$ reduction). However, the standard deviation and reduced iterations could be relatively greater when large initial differences exist among the cells. Thus, the computational cost could be higher and more volatile (due to a higher standard deviation) in complex scenarios, such as could be encountered in actual atmospheric simulations. 

We use the MPI library in the one-core CPU base case experiments instead of disabling it because it is simpler to configure, and the differences are negligible. Specifically, in our tests, disabling the MPI library only reduces the execution time by 0.2\%. This is because MPI is only used for initialization and measuring the execution time.

\begin{figure}[!t]
  \centering
\includegraphics[width=\linewidth]{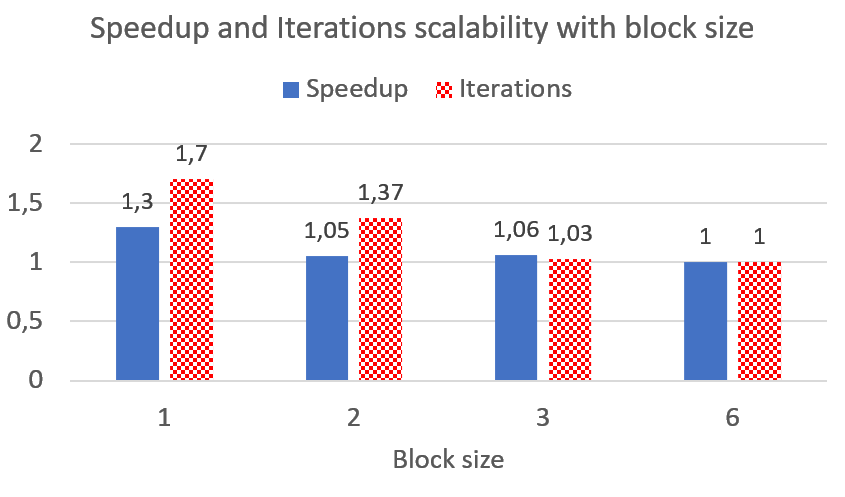}
      \caption{Speedups and iterations of GPU linear solver configurations. The speedup and iterations are normalized over the worst case, which corresponds to Block-cells (N) and Block-cells (3), respectively. The  Block-cells (1), (2),(3), and (N) implementations correspond to block sizes 1,2,3 and 6. The results are averaged over 720 time steps and are configured with 10,000 cells and \emph{realistic} conditions.}
  \label{speedupIterationsvsBlocksize}
\end{figure}

Figure \ref{speedupIterationsvsBlocksize} shows that the speedup decreases linearly as the block size increases, with Block-cells (1) demonstrating the highest performance. This suggests that solving cells individually is faster compared with grouping them. However, despite achieving a reduction of approximately 70\% in the number of solver iterations, the speedup of Block-cells (1) is not as high as expected. This discrepancy implies that there are additional benefits associated with other configurations. We attribute this to using the same solver parameters across multiple cells, which reduces data requirements and subsequently lowers memory usage compared with Block-cells (1).

The figure also reveals a linear decrease in the number of solver iterations with increasing block size, as anticipated due to the increased complexity of the system with more cells involved. Similarly, the speedup exhibits a comparable trend, except for block size 2, where the speedup closely mirrors that of block size 3, despite a significant reduction in iterations between the two cases. This discrepancy suggests another metric influences the speedup besides the iterations, although we have not identified a clear metric in the NVVP report. Further investigation into this behavior may be warranted in the future.

\begin{figure}[!t]
  \centering
\includegraphics[width=\linewidth]{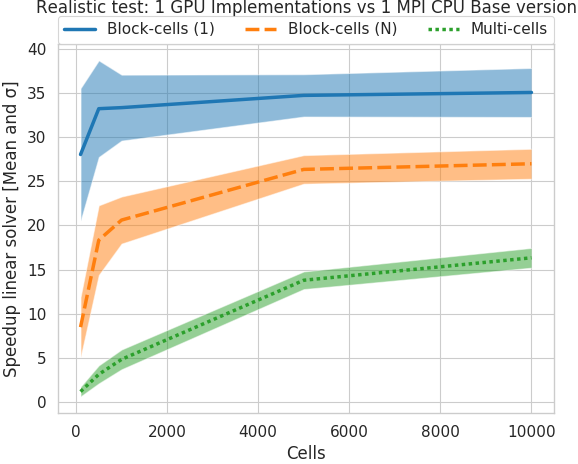}
      \caption{Speedups of GPU linear solver implementations compared to One-cell version executed by one CPU core. The implementations presented are Multi-cells (dotted line), Block-cells (N) (dashed line), Block-cells (1) (continuous line). The speedups are averaged over 720 time steps (as explained in Section \ref{Test environment}). The area covering the speedups is the standard deviation.}
  \label{speedup_linear_solver}
\end{figure}

Figure \ref{speedup_linear_solver} illustrates the speedups of Multi-cells, Block-cells (N), and Block-cells (1) against the CPU--based One-cell implementation using a single core. The standard deviation depicted in Figure \ref{speedup_linear_solver} arises from the inherent differences between the BCG and KLU linear solvers. For instance, BCG is an iterative solver, and the sequence of floating point operations differs between the CPU and GPU, introducing variability into the comparison. The Multi-cells approach exhibits the smallest standard deviation, whereas those for Block-cells (N) and Block-cells (1) are larger. This discrepancy stems from configuring the remainder of the ODE solver code following the Multi-cells approach. Specifically, Block-cells (1) and Block-cells (N) solve the system differently from the ODE solver they are embedded in, resulting in divergent results and increased variance. Block-cells (1) displays more variability than Block-cells (N) because they diverge more significantly from Multi-cells, as the Block-cells (N) approach solves some cells as a single system. 

All the analyzed implementations result in speedups over the single-core CPU implementation. Specifically, Block-cells (N) achieves a 27$\times$ speedup, compared with a 17$\times$ speedup from Multi-cells, indicating an improvement in computational time associated with data transfers between GPUs. Block-cells (1) is the fastest implementation with a 35$\times$ speedup.

Introducing a preconditioner to the linear solver could potentially alter the associated speedups. If the preconditioner effectively reduces the number of iterations required, this reduces more iterations on implementations with higher iteration counts. Consequently, the speedup of Block-cells (N) may approach that of Block-cells (1), given that Block-cells (N) typically entails more iterations than Block-cells (1), as demonstrated in Figure \ref{speedup_iterations_realistic_ideal}. However, a good scaling is not compromised independently of the use or not of a preconditioner. In any case, pursuing an effective preconditioner represents a promising avenue for optimization in future research endeavors.

\begin{figure}[!t]
  \centering
  \includegraphics[width=\linewidth]{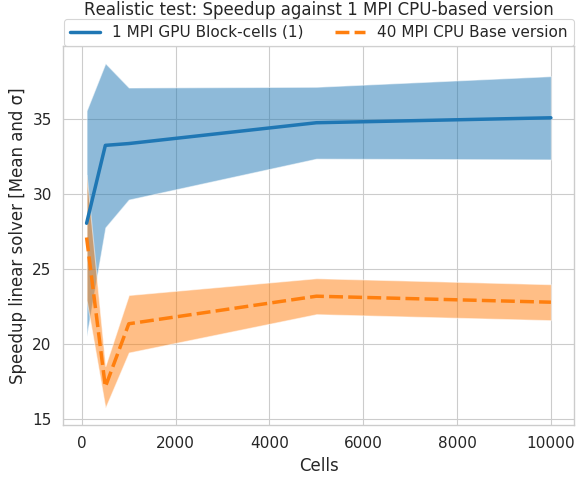}
      \caption{Speedups of GPU linear solver using 1 MPI core (continuous line) and CPU solver using 40 MPI cores (dashed line), both against the base CPU version using 1 MPI core. The speedups are averaged over 720 time steps. The area covering the speedups is the standard deviation.}
  \label{speedup_linear_solver_mpi}
\end{figure}

Figure \ref{speedup_linear_solver_mpi} shows that Block-cells (1) achieves a greater speedup than the 40 cores MPI implementation (35$\times$ vs. 23$\times$). This highlights the performance improvements made possible through the efficient use of the GPU. Moreover, these results can be expected to improve by a factor of 4 when using the 4 GPUs available in the node since no communication is required among the GPUs. We will explore this possibility in future work.

Finally, we inspect the time execution consumed on moving data between GPU and CPU of the best configuration, Block-cells (1). For 10,000 cells, this time on data movement takes 36\% of the total time execution, which is significant. Data transfers cannot be reduced as the implementation already transfers a minimal amount of information, that is, the input and output concentrations. Alternatively, we can augment computational operations by incorporating more code from CVODE into the GPU kernel. However, this approach may require additional data transfers because CVODE uses numerous auxiliary variables. Thus, to offset these transfers, we must translate a substantial portion of the computational workload to the GPU approach, such as developing a GPU--based Block-cells (1) version of CVODE. This option requires further exploration in future investigations.

\section{Conclusions} \label{conclusions}

The main goal of this work has been to achieve an optimized distribution of the computational load of a chemical solver for use on GPUs. We used the CAMP framework to benchmark our developments. The standard approach in CAMP, as in most atmospheric packages, is to compute the chemical state of each cell separately. We refer to this implementation as One-cell. As in the non-parallelized CPU--based solver, these cells are computed one-by-one sequentially. In contrast, the Multi-cells approach is based on grouping the cells of the model into a single system and solving them simultaneously.

Compared with the single-threaded (sequential; not parallelized) solver used by CAMP, Multi-cells has the advantage of requiring significantly fewer solver iterations, as the total number of iterations required for the One-cell approach is the sum of the individual number of iterations needed for each cell \cite{Guzman2021}. This work shows that the Multi-cells performance strongly depends on the initial conditions. The speedup can be as low as 1$\times$ under idealized conditions where all cells have the same initial conditions (i.e., same initial concentrations, temperature, pressure, etc). Nevertheless, the speedup can be as much as 6$\times$ when the initial conditions vary among cells. Therefore, we expect Multi-cells to outperform the single-threaded One-cell approach in CAMP under realistic conditions.

We also evaluated the GPU implementation of the BCG linear solver for the One-cell and Multi-cells cases against the CPU--based KLU solver using the One-cell approach. For the BCG One-cell configuration, the speedup is less than 1$\times$, as there are CPU-executed instructions between the execution of each kernel, preventing the cells from being computed in parallel. The Multi-cells approach gets around this problem by computing cells in parallel, resulting in up to a 17$\times$ speedup for 10,000 cells. However, the Multi-cells implementation allows the load of an individual cell to be divided between two thread blocks, requiring communication between blocks that can account for more than 50\% of the execution time.

To improve the performance of the Multi-cells approach, we proposed the novel Block-cells strategy to avoid communication across thread blocks. The Block-cells approach ensures that each cell is computed within a single block, avoiding communication and synchronization across thread blocks. Block-cells can be configured differently depending on the number of cells assigned per block. This paper compares configuring one cell per block versus using the maximum number of cells per block. We call these configurations Block-cells (1) and Block-cells (N), respectively. Configurations with multiple cells per block but less than the maximum possible result in performance between these two extremes.

We find that both configurations increase the overall memory efficiency by 12\% relative to the Multi-cells approach, indicating improved use of hardware resources. We also show that Block-cells (1) has 15\% more kernel occupancy than Block-cells (N), meaning it has more active threads per warp. Moreover, we show that Block-cells (1) requires $\sim$80\% fewer solver iterations than Block-cells (N). This leads to different performance improvements for Block-cells (N) and Block-cells (1), which attain up to 27$\times$ and 35$\times$ speedups, respectively, relative to the CPU--based One-cell configuration.

We also compare the speedup obtained over an equivalent MPI implementation, which uses the maximum number of physical cores available on a node (40). In this MPI implementation, we emulate an actual ESM experiment and solve the number of cells in each process equal to the total cells divided by the number of processes. Block-cells (1) is up to 50\% faster than this MPI implementation. This highlights the advantage of the GPU--based Multi-cells approach over traditional CPU--based approaches for the hardware specifications used in this study.

In summary, the new Block-cells strategy improves upon the previously developed GPU--based Multi-cells approach and a traditional CPU--based parallel implementation using MPI. Moreover, we present evidence that the Block-cells approach can be an excellent alternative to other GPU--based deployments, in which the workload of a cell is handled by a single thread, not by a thread block. It should also be noted that the Block-cells approach can be applied to the rest of the CAMP ODE solver algorithm, which now represents $\sim$95\% of the total solving time after applying the Block-cells strategy to the linear solver.

Thus, in future work, we will apply the Block-cells strategy to the rest of the BDF algorithm. We expect this to improve the performance relative to other GPU--based chemical solvers, as the complete algorithm will have been ported to GPUs. More specifically, we expect to obtain a speedup similar to that found for the linear solver, as the BDF algorithm is conceptually similar to the BCG algorithm: both are iterative algorithms based on similar algebraic operations (i.e., vector multiplication or vector reduction to a variable). However, the BDF algorithm is more complex than the linear solver and may present unique challenges. We can also use the CPU and GPU solvers in a hybrid implementation, where OpenMP may be a promising option to improve the CPU computation. In future work, we will consider these options and investigate the viability of the Block-cells approach for a GPU--based version of this algorithm.

\section*{Code availability}
The code used in this study is available at the CAMP GitHub repository 
at \url{https://github.com/open-atmos/camp} (last access: 15 Mar 2024) and at Mendeley Data \cite{guzman_ruiz_camp_nodate}.

\section*{Declaration of competing interest}
The authors declare that they have no known competing financial interests or personal relationships that could have appeared
to influence the work reported in this paper.

\section*{Acknowledgment}
This work was partially supported by funding from the Ministerio de Ciencia, Innovación y Universidades as part of the BROWNING project (RTI2018-099894-BI00 funded by MCIN AEI/10.13039/
501100011033) and part of the CAROL project (PID2020-113614RBC21 funded by MCIN AEI/10.13039/501100011033), the Generalitat de Catalunya (2021-SGR-00574, 2021-SGR-01550 and 2021-SGR00785) and the AXA Research Fund through the AXA Chair on Sand and Dust Storms established at the Barcelona Supercomputing Center (BSC). Matthew Dawson has received funding from the European
Union’s Horizon 2020 research and innovation program under Marie Skłodowska-Curie grant agreement no. 747048. Mario Acosta has received funding from the National Research Agency through OEMES (PID2020-116324RA-I00). This paper expresses the opinions of the authors and not necessarily those of the funding commissions.

\bibliographystyle{elsarticle-num}
\bibliography{zotero}

\end{document}